\documentstyle[12pt]{article}
\begin{document}

\rightline{FTUV 94-44}
\rightline{IFIC 94-39}

\begin{center}
{\Large{\bf Non-commutative geometry and covariance: from the quantum plane
to quantum tensors$^*$}}
\end{center}

\vspace{2\baselineskip}

\begin{center}
{\large{J.A. de Azc\'{a}rraga$^{\dagger}$, P.P. Kulish\footnote{On
leave of absence from the St.Petersburg's Branch of the Steklov Mathematical
Institute of the Russian Academy of Sciences.

$^*$Lecture delivered (by J.A.) at the 3rd Colloquium
`Quantum groups and Physics', Prague, June 1994. To appear in the proceedings
(Czechoslovak J. Phys.). }$^{\dagger}$
and F. Rodenas$^{\dagger \; \ddagger}$}}
\end{center}
\vspace{0,5cm}
\addtocounter{footnote}{-1}

\begin{center}
{\small{\it $ \dagger \quad$ Departamento de F\'{\i}sica Te\'{o}rica and
IFIC,\\
Centro Mixto Universidad de Valencia-CSIC\\
E-46100-Burjassot (Valencia), Spain.}}
\end{center}

\begin{center}
{\small{\it $ \ddagger \quad$ Departamento de Matem\'atica Aplicada,\\
 Universidad Polit\'ecnica de Valencia\\
E-46071 Valencia, Spain.}}
\end{center}
\vspace{1cm}

\begin{abstract}
Reflection and braid equations for rank two $q$-tensors are derived from the
covariance properties of quantum vectors by using the $R$-matrix formalism.
\end{abstract}

\section{Introduction}

Quantum groups may be looked at in various ways. From a mathematical point of
view, they may be introduced by making emphasis on their
$q$-deformed enveloping algebra aspects \cite{DRINFELD,JIMBO} or by making
emphasis in the $R$-matrix formalism that describe the deformed group algebra
\cite{FRT}. A point of view which is particularly useful in possible
physical applications
is to look at quantum groups as symmetries of {\it quantum spaces}
\cite{MANIN,WZ}. The simplest example of this approach is constituted by
the well known  quantum plane $C_{q}^{2}$, or associative
{\it algebra} (a $q$-plane is not a manifold) generated by two elements
$(x,y)=X$ (a $q$-two-vector) subjected to the commutation property \cite{MANIN}
\begin{equation}\label{1a}
xy = qyx \quad .
\end{equation}
\noindent
The commutation relation (\ref{1a}) can also be expressed  by using the
$q$-symplectic metric $\epsilon^q$
\begin{equation}\label{epsilon1}
\epsilon^q = \left( \begin{array}{cc}
                       0 & q^{-1/2} \\
                       -q^{1/2} & 0
                     \end{array} \right)  \quad , \qquad
( \epsilon^q )^2=-I
\end{equation}
\noindent
by the  equation
\begin{equation}\label{epsilon2}
X^t \epsilon^q X =0 \quad,\qquad \epsilon^q_{ij} X_iX_j=0
\end{equation}
\noindent
which reflects that the $q$-symplectic norm of a $q$-two-vector  vanishes.

It is also possible to introduce a pair of (odd) variables $(\xi,\eta)=\Omega$
(an odd $q$-two-vector)  satisfying
\begin{equation}\label{odd}
\xi \eta = - \frac{1}{q} \eta \xi \quad ,\qquad \xi^2 =0= \eta^2 \quad.
\end{equation}
\noindent
If it is required that, after the transformations $X'=TX$,
$\Omega'=T \Omega$ the new entities
$(x',y')$, $(\xi',\eta')$ satisfy also (\ref{1a}), (\ref{odd}),
it is found that the commutation properties of the elements of $T$
\begin{equation}\label{1b}
T=
\left(
\begin{array}{ll}
a & b\\
c & d
\end{array}
\right)
\end{equation}

\noindent
are completely determined. These are the well known relations (the entries
of $T$ commute with those of  $X$ and $\Omega$)
\begin{equation}\label{1c}
\begin{array}{lll}
ab=qba \quad , & \quad ac=qca \quad , & \quad bd=qdb \quad,\\
cd=qdc \quad , & \quad [a, d] = \lambda cb \quad , & \quad [b, c]=0 \quad,
\end{array}
\end{equation}

\noindent
$(\lambda \equiv q - q^{-1})$
which constitute a presentation of the $GL_{q}(2,C)$ algebra generated by
$(a,b,c,d)$. For $q$=1, $(x,y)$ commute and $(\xi,\eta)$ anticommute.
In a non-commutative differential calculus this second set of variables
are identified with the differentials (\cite{WZ}) of $(x,y)$.
Here we shall consider the quantum plane (\ref{1a}) only
as the representation (co-module) space of the $GL_q(2,C)$ quantum group
(\ref{1c}).
In terms of the $R$-matrix formalism \cite{FRT}, eqs. (\ref{1a}) and
(\ref{1c}) may be written as (see eq. (\ref{1i}) below)
\begin{equation}\label{1d}
R_{12} X_{1} X_{2} =  q X_{2} X_{1}
\quad \Longleftrightarrow \quad
R_{21}^{-1} X_{1} X_{2} =  q^{-1} X_{2} X_{1}\quad,
\end{equation}
\begin{equation}\label{1e}
R_{12} T_{1} T_{2} = T_{2} T_{1} R_{12}
\quad \Longleftrightarrow \quad
R_{21}^{-1} T_{1} T_{2} = T_{2} T_{1} R_{21}^{-1}\quad,
\end{equation}

\noindent
where the standard  notation $T_{1} = T \otimes {\bf 1} \; , \; T_{2} = {\bf 1}
\otimes T \quad (T_{1 ij,kl} = T_{ik} \delta_{jl}\; , \; T_{2ij,kl}=
\delta_{ik} T_{jl},\; i,j,k,l= 1,2)$ has been used, and $X_{1} X_{2}$ and
$X_{2} X_{1}$ are, respectively, the four-vectors $(xx,xy,yx,yy)$ and
$(xx,yx,xy,yy)$.
Both relations in (\ref{1d}) (and in (\ref{1e})) are equivalent.
This is easy to see  by using
the permutation operator ${\cal P}$ which gives
$({\cal P}R{\cal P})_{ij,kl}=R_{ji,lk}$
$({\cal P}R_{12}{\cal P}=R_{21})$  and $({\cal P}X_1X_2)_{ij}=(X_1X_2)_{ji}$
{\it i.e.}, ${\cal P}X_1X_2=X_2X_1$.
In this matrix notation it is obvious  that (\ref{1e}) is
consistent with
the requirement of  invariance of (\ref{1d}) under the transformation
$X'=TX$,
$R_{12} X_{1}' X_{2}' = q X_{2}' X_{1}'$. Since the elements of $X$ commute
with the entries of $T$, we obtain
\begin{equation}\label{1f}
\begin{array}{l}
R_{12} X_{1}' X_{2}' = R_{12} (T_{1} X_{1})
(T_{2} X_{2})=
R_{12} T_{1} T_{2} X_{1} X_{2} \\
\, \\
= T_{2} T_{1} R_{12}
X_{1} X_{2} =q T_{2} T_{1} X_{2} X_{1} = q X_{2}' X_{1}'\quad
\end{array}
\end{equation}

\noindent
using (\ref{1e}), and the invariance of  (\ref{1d}) follows: the preservation
of (\ref{1d})
under the `$q$-symmetry'
transformation requires (\ref{1e}). In components, (\ref{1d})  reads
\begin{equation}\label{1g}
R_{ij,kl} X_{k} X_{l} = q X_{j} X_{i} \quad ,\qquad
\hat{R}_{ij,kl} X_{k} X_{l} = q X_{i} X_{j} \quad,
\end{equation}

\noindent
where $R_{12}$ and  ${\cal P}$ ($\hat{R} = {\cal P}R$,  $\hat{R}_{ij,kl}=
R_{ji,kl})$ are given by

\begin{equation}\label{1i}
R=
\left[
\begin{array}{llll}
q & \, & \, & \, \\
\, & 1 & 0 & \, \\
\, & \lambda & 1 & \, \\
\, & \, & \, & q
\end{array}
\right]
\quad , \quad
{\cal P}=
\left[
\begin{array}{llll}
1 & \, & \, & \, \\
\, & 0 & 1 & \, \\
\, & 1 & 0 & \, \\
\, & \, & \, & 1
\end{array}
\right]
\quad .
\end{equation}

Although the indices in all previous expressions take the values $1, 2$,
the $R$-matrix form of the basic expressions
(\ref{1d}) and (\ref{1e}) makes it clear how to
generalize them to $GL_{q}(n, C)$; all that is needed is the appropriate
$n^{2} \times n^{2}$  $R$-matrix, which is given by \cite{FRT}
\begin{equation}\label{1j}
R_{ij,kl} = \delta_{ik} \delta_{jl} (1 + \delta_{ij} (q-1)) + \lambda
\delta_{il} \delta_{jk} \theta (i-j) \quad i,j...=1...n
\end{equation}
$$
\theta (i-j) = \left\{ \begin{array}{l}
                      0 \quad i \leq j \\
                      1  \quad i > j
                      \end{array} \right.  \quad.
$$

\noindent
With it, the relations defining the `quantum hyperplane'

\begin{equation}\label{1k}
X=(x_{1},...,x_{n})\quad , \quad
x_{i} x_{j} = q x_{j} x_{i} \quad (i<j)\; i,j=1...n
\end{equation}

\noindent
are again expressed by (\ref{1d}) and preserved under $GL_{q}(n, C)$
because of (\ref{1e}).

All this is, of course,  well known. In this report we exhibit  how to extend
these $q$-vector constructions to higher rank quantum tensors
(see also \cite{K-SK,K-SA}).
In particular, we shall consider the simplest example of $q$-twistors
constructed from $q$-two-vectors (spinors) (\ref{1a}),
(\ref{epsilon2}), (\ref{1d}) and
the  application to $q$-Minkowski space algebras \cite{AKR}.

\section{Other covariant objects. Quantum twistors}

Consider two isomorphic objects $X$ and $Z$, and their hermitian conjugates
$X^{\dagger}$ and $Z^{\dagger}$, transforming under the coaction of two
different quantum groups $T$ and $T^{\dagger}$ by
\begin{equation}\label{2a}
\begin{array}{ll}
X'=T X \quad \quad ,& \quad \quad X'^{\dagger} = X^{\dagger} T^{\dagger}
\quad ,\\
Z'=T Z  \quad \quad ,& \quad \quad Z'^{\dagger} = Z^{\dagger} T^{\dagger}
\quad ,
\end{array}
\end{equation}

\noindent
For instance, in the classical $SL(2,C)$
case there are two fundamental representations,
$D^{\frac{1}{2},0}$ and $D^{0,\frac{1}{2}}$, realized by complex unimodular
matrices $A$ and $(A^{-1})^{\dagger}$. In the quantum case this corresponds to
taking two copies $T$ and $\tilde{T}$ of $SL_{q}(2,C)$,
with the obvious `reality' condition
added, $\tilde{T}^{-1} =T^{\dagger}$. In the $q$-case one has to add the
commutation relations between elements of $T$ and $T^{\dagger}$.

The commutation relations in the general situation involve
four $R$-matrices $R^{(i)}$, $i=1,...,4$,

\begin{equation}\label{2b}
\begin{array}{l}
R^{(1)} T_{1} T_{2} =  T_{2} T_{1} R^{(1)}\quad ,\\
\, \\
T_{1}^{\dagger} R^{(2)} T_{2} =  T_{2} R^{(2)} T_{1}^{\dagger}\quad ,\\
\, \\
T_{2}^{\dagger} R^{(3)} T_{1} =  T_{1} R^{(3)} T_{2}^{\dagger}\quad, \\
\, \\
R^{(4)} T_{1}^{\dagger} T_{2}^{\dagger} =  T_{2}^{\dagger} T_{1}^{\dagger}
R^{(4)}\quad .
\end{array}
\end{equation}

\noindent
The consistency of these equations requires
$R^{(2)} ={\cal P} R^{(3)} {\cal P}=R^{(3) \,\dagger}$ and
$R^{(4)}=R^{(1) \dagger}$ or
$R^{(4)}=({\cal P}R^{(1)\;-1}{\cal P})^{\dagger}$.
Notice that $R^{(1)}$ and $R^{(4)}$ are the
$R$-matrices of two quantum groups $T$ and $T^{\dagger}$
related by a $*$-operation (and that $R^{(1)}$, {\it e.g.}, may be taken as
$R_{12}$ or $R_{21}^{-1}$). In contrast,
$R^{(2)}$ (and hence $R^{(3)}$) is a matrix defining how the elements of both
quantum groups
commute and accordingly it is not a priori fixed. In general,
one could  introduce instead of $T^{\dagger}$ another matrix $S$;
the $q$-matrices
$S$ and $T$ need not even have the same dimension. If, say, $T$ and $S$ are
$n \times n$ and $m \times m$ matrices, $S_{1}$ and $T_{2}$ in the
second equation of (\ref{2b})
would be  $S_{1}^{\dagger}= (S^{\dagger}\otimes {\bf 1}_{n})$ and $T_{2}=
({\bf 1}_{m} \otimes T)$ and  $R^{(2)}$ would be  an $(m \times n) \times
(m \times n)$ matrix. Similarly, in the third equation $S_{2}^{\dagger}
= ({\bf 1}_{n} \otimes S^{\dagger})$, $T_{1} = (T \otimes
{\bf 1}_{m})$ and $R^{(3)}$ would be an $(n \times m) \times (n \times m)$
matrix,
while $R^{(4)}$  would be an $m^2 \times m^2$  matrix.

The form of the eqs. (\ref{2b}) is the result of the equations which express
the
commutation relations among the components of the vectors $X,Z$. Since in
principle  $T$ and $T^{\dagger}$ do not commute, we have to allow for
possibly
non-trivial commutation relations among the components of $X$ and
$Z^{\dagger}$.
Thus, the set of commutation relations left invariant  is given by

\begin{equation}\label{2c}
\begin{array}{l}
R^{(1)} X_{1} X_{2} =  \kappa_{1} X_{2} X_{1}\quad ,\\
\, \\
Z_{1}^{\dagger} R^{(2)} X_{2} =  X_{2} Z_{1}^{\dagger}\quad ,\\
\, \\
Z_{2}^{\dagger} R^{(3)} X_{1}  = X_{1} Z_{2}^{\dagger}\quad ,\\
\, \\
\kappa_{2}Z_{1}^{\dagger} Z_{2}^{\dagger} = Z_{2}^{\dagger} Z_{1}^{\dagger}
R^{(4)}\quad ,
\end{array}
\end{equation}

\noindent
where $\kappa_{1}$ and $\kappa_{2}$ are
appropriate eigenvalues of the $R$-matrices.
The invariance of the first and last equations is
proven as in Sec.1 and the others similarly. For instance, for the second
equation we check that

$$
Z_{1}^{'\dagger} R^{(2)} X_{2}' =
(Z_{1}^{\dagger} T_{1}^{\dagger}) R^{(2)} (T_{2} X_{2})= Z_{1}^{\dagger}
T_{2} R^{(2)} T_{1}^{\dagger} X_{2}
$$

\begin{equation}\label{2d}
=T_{2} Z_{1}^{\dagger} R^{(2)} X_{2} T_{1}^{\dagger} = T_{2} X_{2}
Z_{1}^{\dagger} T_{1}^{\dagger} = X_{2}' Z_{1}^{'\dagger}
\end{equation}

\noindent
using the second equations in (\ref{2b}) and (\ref{2c}), respectively, in the
second and fourth equalities. In particular, if $R^{(2)} = I=
R^{(3)},$ both quantum groups are independent (commuting), and this is
reflected in
the fact that the components of $X$ and $Z^{\dagger}$ commute.

Let us use the above construction to introduce another  covariant object
which generalizes (with some restrictions)   the concept of twistor
to the $q$-deformed case.
Let $X$ and $Z^{\dagger}$  satisfy the previous set of commutation
relations. In particular, $X$ and $Z^{\dagger}$ may be, for instance,
$q$-two-vectors ($q$-spinors),
of $SL_{q} (2, C)$;
this case will be analyzed in more detail below. Tensoring two $q$-vectors we
introduce the object
\begin{equation}\label{2e}
K \equiv X Z^{\dagger} \qquad (K_{ij}=X_iZ^{\dagger}_j)\quad.
\end{equation}

\noindent
Then, the transformation of $K$ induced by (\ref{2a})  is
\begin{equation}\label{2f}
\varphi:\;K \longmapsto K'= TK T^{\dagger} \qquad
(K'_{ij}=T_{im}K_{mn}T^{\dagger}_{nj})\quad.
\end{equation}

\noindent
The entries of $K$ are, of course, non-commuting. We shall see that these
commutation relations can be  expressed in a closed, elegant
and compact equation
which permits to extract the algebra generated by the entries of $K$
without considering its  explicit realization in terms of the components of $X$
and
$Z^{\dagger}$.
Using the above relations we may now derive the equation describing the
commutation relations which define the algebra generated by the entries of
$K$. With $K_{1} = X_{1} Z_{1}^{\dagger}$
$(K_{1\;ij,kl}=(K \otimes {\bf 1} )_{ij,kl}=X_iZ^{\dagger}_k \delta_{jl})$
and $K_{2}= X_{2} Z_{2}^{\dagger}$
$(K_{2\;ij,kl}=( {\bf 1} \otimes K)_{ij,kl}= \delta_{ik}X_jZ_l^{\dagger})$,
we find using (\ref{2c}) that

\begin{equation}\label{2fa}
\begin{array}{l}
R^{(1)} K_{1} R^{(2)} K_{2} = R^{(1)} X_{1} Z_{1}^{\dagger} R^{(2)} X_{2}
Z_{2}^{\dagger}
=R^{(1)} X_{1} X_{2} Z_{1}^{\dagger} Z_{2}^{\dagger} \\

\qquad = (\kappa_{1}/\kappa_{2} ) X_{2} X_{1}
Z_{2}^{\dagger} Z_{1}^{\dagger} R^{(4)}=
(\kappa_{1}/\kappa_{2})
X_{2} Z_{2}^{\dagger} R^{(3)} X_{1} Z_{1}^{\dagger} R^{(4)}\quad .
\end{array}
\end{equation}

\noindent
Hence, the commuting properties of the quantum twistor are given by

\begin{equation}\label{2g}
R^{(1)} K_{1} R^{(2)} K_{2} = (\kappa_{1}/\kappa_{2})
K_{2} R^{(3)} K_{1} R^{(4)}
\end{equation}

\noindent
Eq. (\ref{2g}) is (with $\kappa_{1}/\kappa_{2} =1$)
nothing else than the reflection equation with no spectral parameter
dependence (see \cite{K-SK,K-SA} and references therein and  \cite{MAJID} in
the
context of braided algebras)
which follows by imposing the invariance of the commuting properties of the
entries of $K$ by the coaction (\ref{2f}). As shown here, eq. (\ref{2g})
also follows from  interpreting $K$ as an object made out of two $q$-`vectors',
in general not necessarily of the same dimension so that in general $K$ is not
a squared matrix.

Let $X$, $Z$ be two $q$-two-vectors (spinors). Then, $K=XX^{\dagger}$
is a (null) {\it quantum twistor}:  as we shall see, its quantum determinant
($det_qK$)  is necessarily zero (as it is as well for $XZ^{\dagger}$).
In contrast,
the $q$-twistor  $K=XZ^{\dagger}+ZX^{\dagger}$ has   $det_qK \neq 0$.

Notice that, in general, there are four possibilities to write (\ref{2g})
(obviously, related in between)  since there are two possibilities for
$R^{(1)}$ and for $R^{(4)}$ in (\ref{2b}) and in (\ref{2c})
(see (\ref{1d}) and  (\ref{1e})).
However, this freedom is reduced when covariant objects $K$
constructed out of four
vectors are considered since covariance requires to introduce commutation
relations  between $Z$ and $X$ and between $Z^{\dagger}$ and $X^{\dagger}$
using $R^{(1)}$ and $R^{(4)}$.
Let us consider the hermitian matrix
\begin{equation}\label{tw1}
K=XZ^{\dagger}+ZX^{\dagger}
\end{equation}
\noindent
($Z$ and $X$ have the same number of components).
To compute the commutation properties of $K$, the complete set of relations
among  $X$, $Z$, $X^{\dagger}$ and $Z^{\dagger}$ are required. Thus, besides
(\ref{2c}), we need to introduce the following set of covariant relations
\begin{equation}\label{tw2}
\begin{array}{l}
R^{(1)} X_{1} Z_{2} =  Z_{2} X_{1}\quad ,\\
\, \\
Z_{1}^{\dagger} R^{(2)} Z_{2} =  Z_{2} Z_{1}^{\dagger}\quad ,\\
\, \\
X_{2}^{\dagger} R^{(3)} X_{1}  = X_{1} X_{2}^{\dagger}\quad ,\\
\, \\
X_{1}^{\dagger} Z_{2}^{\dagger} = Z_{2}^{\dagger} X_{1}^{\dagger}
R^{(4)}\quad ,
\end{array}
\end{equation}
\noindent
the structure of which is again dictated from (\ref{2b}) by covariance.
{}From the first and  the last eqs. in (\ref{tw2}) we obtain
(supposing $R^{(1)}$ real)
\begin{equation}\label{tw3}
R^{(4)}=(R^{(1)})^t
\end{equation}
\noindent
which implies that  the eigenvalues are equal, $\kappa_1= \kappa_2$. Then,
the commutation relation for the entries of $K$ (in matrix form)
are easily computed using (\ref{2c}) and  (\ref{tw2})
\begin{equation}\label{tw4}
\begin{array}{rcl}
R^{(1)}K_1R^{(2)}K_2 &=& R^{(1)} (X_1Z_1^{\dagger}+Z_1X_1^{\dagger})
R^{(2)}(X_2Z_2^{\dagger}+Z_2X_2^{\dagger}) \\
\; &=& (X_2Z_2^{\dagger}+Z_2X_2^{\dagger})R^{(3)}
(X_1Z_1^{\dagger}+Z_1X_1^{\dagger})R^{(4)} \\
\; &=& K_2R^{(3)}K_1R^{(4)}
\end{array}
\end{equation}
\noindent
where $R^{(1)}=R^{(4)\,t}={\cal P}R^{(4)}{\cal P}$ and
$R^{(2)}={\cal P}R^{(3)}{\cal P}$ and Hecke's condition
for the $R$-matrix has been used. Now, we have only one reflection equation
for $K$, since the two possibilities for $R^{(1)}$
produce two equations which are identical after a similarity transformation
with
${\cal P}$.

Notice that $K$ in (\ref{tw1}) is constructed from two parts, each one
of them satisfying the
same algebra relations (\ref{tw4}):
\begin{equation}\label{br1}
\quad K=K^{(1)}+K^{(2)}\quad,\quad
K^{(1)}=XZ^{\dagger} \quad , \quad K^{(2)}=ZX^{\dagger} \quad.
\end{equation}
\noindent
These two pieces have  specific commutation properties among themselves.
Indeed,
the (mixed) commutation relations (\ref{tw2}) lead to the following
non-commuting property
between the matrices $K^{(1)}$ and $K^{(2)}$
(non-symmetric under the interchange of $K^{(1)}$ and $K^{(2)}$)
\begin{equation}\label{br2}
R^{(1)}K^{(1)}_1R^{(2)}K^{(2)}_2=
K^{(2)}_2R^{(3)}K^{(1)}_1({\cal P}R^{(4)}{\cal P})^{-1} \quad.
\end{equation}
\noindent
Here $R^{(4)}=R^{(1)\;t}$ and the two possibilities for $R^{(1)}$
produce two different equations for  $K^{(1)}$ and $K^{(2)}$
which transform into each other by exchanging (1)$ \leftrightarrow $(2)
in $K^{(i)}$. Both had to be possible since $K=K^{(1)}+K^{(2)}$
is symmetric under this exchange.
Equation (\ref{br2}), here obtained  from the commutation relations
(\ref{tw2}), is known as `braiding equation' \cite{MAJID}.
The commutation properties among the elements of $K^{(1)}$ and $K^{(2)}$
are such that the sum of two objects satisfying
(\ref{tw4}) verifies  also the same relation. Within this terminology,
the `mixed' eqs.
(\ref{tw2}) are the braiding relations for $q$-vectors.

{}From now on, we shall restrict ourselves to the two-dimensional case which
will
be  useful in the application to $q$-Minkowski space \cite{AKR}. We shall
start by discussing the

\vspace{1\baselineskip}

\noindent
{\bf $q$-determinant of $K$:}

\vspace{1\baselineskip}
\indent
Let  the quantum group matrices $T$ and $T^{\dagger}$ be 2$\times$2 matrices.
There exists an invariant quadratic element from $K$, the
$q$-determinant of $K$. It is defined by \cite{K-SK}
\begin{equation}\label{det1}
det_qK \, P_- \equiv P_- K_1 \hat{R}^{(3)} K_1 P_-
\end{equation}
\noindent
where $\hat{R}^{(3)}={\cal P}R^{(3)}$ and $P_-$ is the $q$-antisymmetrizer of
the $R$-matrix corresponding to the quantum groups $T$ and $T^{\dagger}$.
The $q$-determinant
of $T$ and $T^{\dagger}$ are given by \cite{FRT}
\begin{equation}\label{det2}
det_qT \, P_- = P_- T_1T_2 \qquad , \qquad det_qT^{\dagger} \, P_- =
T_2^{\dagger}
T_1^{\dagger}P_-
\end{equation}
\noindent
and  the projector $P_-$ can be expressed in terms of the $q$-epsilon
tensor (\ref{epsilon1})
\begin{equation}\label{det4}
P_{-\; ij,kl}=[2]^{-1}\epsilon^q_{ij}\epsilon^q_{kl}\quad,\qquad
P_-= \frac{1}{[2]} \left[ \begin{array}{cccc}
                        0  &  0  &  0  &  0 \\
                        0  &q^{-1}& -1 &  0 \\
                        0   &  -1 &  q &   0 \\
                        0   &  0  &  0  &  0
                \end{array}   \right] \quad.
\end{equation}
\noindent
When $(det_qT)(det_qT^{\dagger})=1$, $det_qK$ is invariant under the coaction
(\ref{2f}). Using the third eq. in (\ref{2b}) and (\ref{det2})
\begin{equation}\label{det3}
\begin{array}{ccl}
det_q(TKT^{\dagger})&=& P_- (T_1K_1T_1^{\dagger})
\hat{R}^{(3)}(T_1K_1T_1^{\dagger})P_- \\
\, &=& P_-T_1T_2K_1\hat{R}^{(3)} K_1T_2^{\dagger}T_1^{\dagger}P_- \\
\, &=& (det_qT) \,(det_qK)\,(det_qT^{\dagger}) \;.
\end{array}
\end{equation}
\noindent
Thus, if $(det_qT)(det_qT^{\dagger})=1$ we obtain that $det_q(TKT^{\dagger})=
det_qK$.
The centrality of $det_qK$ requires some YBE-like conditions on the
$R^{(i)}$ ($i=1,2,3,4$) matrices in (\ref{2g}).

Using the definition (\ref{det1}) and the $R$-matrix property
$\hat{R}^{(3)}_{ab,cd}=R^{(3)}_{ba,cd}$, we can compute explicitly the
$q$-determinant of $K$ in the following realizations

\vspace{1\baselineskip}
\noindent
{\bf 1.} for the matrix $K=XZ^{\dagger}$
(and hence for the $q$-twistor $K=XX^{\dagger}$)
\begin{equation}\label{det5}
\begin{array}{ccl}
(det_qK)P_{-\; ij,kl}&=& P_{-\; ij,ab}K_{ac}\hat{R}^{(3)}_{cb,mn}
K_{mp}P_{-\;pn,kl} \\
\, & \propto & \epsilon^q_{ij} \epsilon^q_{ab} X_a Z^{\dagger}_c
R^{(3)}_{bc,mn}
X_mZ^{\dagger}_p \epsilon^q_{pn}\epsilon^q_{kl} \\
\, &=&\epsilon^q_{ij} \epsilon^q_{ab} X_a X_b Z^{\dagger}_n
Z^{\dagger}_p \epsilon^q_{pn}\epsilon^q_{kl} \\
\, &=& \epsilon^q_{ij}(X^t\epsilon^qX)(Z^t\epsilon^qZ)^{\dagger}\epsilon^q_{kl}
=0
\end{array}
\end{equation}
\noindent
since $(X^t\epsilon^qX)=0=(Z^t\epsilon^qZ)$. This reflects the well-known
fact in non deformed twistor theory that twistors constructed out of
two spinors determine  null length vectors;

\vspace{1\baselineskip}
\noindent
{\bf 2.} for the $q$-twistor $K=XZ^{\dagger}+ ZX^{\dagger}$, a similar calculus
to the previous one gives
\begin{equation}\label{det6}
\begin{array}{ccl}
(det_qK)P_{-\; ij,kl}& \propto &
\epsilon^q_{ij} \epsilon^q_{ab} (X_a Z^{\dagger}_c
+Z_a X^{\dagger}_c) R^{(3)}_{bc,mn}
(X_mZ^{\dagger}_p+Z_m X^{\dagger}_p) \epsilon^q_{pn}\epsilon^q_{kl} \\
\, &=& \epsilon^q_{ij}[(X^t\epsilon^q Z)(X^t\epsilon^q Z )^{\dagger}
+(Z^t\epsilon^q X)(Z^t\epsilon^q X )^{\dagger}]\epsilon^q_{kl} \;
\neq 0
\end{array}
\end{equation}
\noindent
then, to get twistors with non-null $q$-determinant we need four spinors
in the definition of $K$ (notice that $X$, $X^{\dagger}$, $Z$ and $Z^{\dagger}$
are algebraically independent  objects).
If  the scalar products
$(X^t\epsilon^q Z)$ and $(Z^t\epsilon^q X )$ are  central elements
in the algebra generated by  $X$, $Z$, $X^{\dagger}$ and $Z^{\dagger}$
the $q$-determinant of $K$ is also central.

\section{An application: $q$-Minkowski space}

The classical construction of a Minkowski vector uses two (dotted and
undotted) spinors,

\begin{equation}\label{3a}
K_{\alpha \dot{\beta}} = \xi_{\alpha} \xi_{\dot{\beta}} =
(\sigma_{\mu} x^{\mu})_{\alpha \dot{\beta}}\quad \alpha, \dot{\beta}=1,2 \quad,
\end{equation}

\noindent
and $K'_{\alpha \dot{\beta}} = A_{\alpha} \, .^{\gamma} K_{\gamma \dot{\delta}}
(\tilde{A}^{-1})^{\dot{\delta}} . _{\dot{\beta}} \; ,$ where $A$ and
$\tilde{A} = (A^{-1})^{\dagger}$ are the two fundamental representations of
$SL(2,C)$. A $q$-deformation of the Lorentz group may be obtained
\cite{WSSW}-\cite{OSWZ}
by replacing $A$ and $\tilde{A}$ by two copies $T$ and $\tilde{T}$ of
$SL_{q} (2,C)$. Applying the pattern described above we now have two   pairs of
$q$-spinors

\begin{equation}\label{3b}
X\rightarrow X' =TX \quad \quad Z\rightarrow Z' =TZ \quad,
\end{equation}

\begin{equation}\label{3c}
X^{\dagger} \rightarrow X'^{\dagger} =
X^{\dagger} \tilde{T}^{-1}\quad \quad
Z^{\dagger} \rightarrow Z'^{\dagger} =
Z^{\dagger} \tilde{T}^{-1}\quad ,
\end{equation}

\noindent
obviously, the reality condition $T^{\dagger}=\tilde{T}^{-1}$
must be considered to have that $(X)^{\dagger}=X^{\dagger}$,
from which we may construct  the following hermitian objects ($q$-twistors)

\begin{equation}\label{3d}
K=XX^{\dagger} \quad \mbox{or} \quad K=XZ^{\dagger}+ZX^{\dagger} \quad ,
\end{equation}
\noindent
and
find their transformation properties.
When the reality condition $T^{\dagger}=\tilde{T}^{-1}$ is imposed, the
coaction
\begin{equation}\label{3e}
K' =TK\tilde{T}^{-1} =TKT^{\dagger} \quad
\end{equation}

\noindent
preserves the hermiticity property of $K$.
Since, by assumption, $T$ and $\tilde{T}$ are $SL_q(2,C)$ matrices, {\it i.e.},

\begin{equation}\label{3f}
R_{12} T_{1} T_{2} = T_{2} T_{1} R_{12} \quad , \quad
R_{12} \tilde{T}_{1} \tilde{T}_{2} = \tilde{T}_{2} \tilde{T}_{1} R_{12}\quad ,
\end{equation}

\noindent
the first and last equations of (\ref{2b})  are fulfilled if

\begin{equation}\label{3g}
\begin{array}{l}
R^{(1)} = R_{12} \quad \mbox{or} \quad R_{21}^{-1} \quad
(\kappa_1=q\quad \mbox{or} \quad q^{-1})\quad, \\
R^{(4)} = R_{21} \quad \mbox{or} \quad R_{12}^{-1}\quad
(\kappa_2=q^{-1}\quad \mbox{or} \quad q)  \quad .
\end{array}
\end{equation}

\noindent
Then, the basic relations which define the non-commutative algebra
generated by the entries of $K$ are given by eq. (\ref{2g}),
which gives the following possibilities
\begin{equation}\label{3h}
R_{12} K_{1} R^{(2)} K_{2} =  K_{2} R^{(3)} K_{1} R_{21} \quad .
\end{equation}
\begin{equation}\label{3i}
R_{12} K_{1} R^{(2)} K_{2} = q^2  K_{2} R^{(3)} K_{1} R_{12}^{-1} \quad .
\end{equation}

\noindent
As we have already discussed the second possibility (\ref{3i}) is not valid for
the twistor with four spinors (second  expression in (\ref{3d})) since
it does not correspond to $R^{(1)}=R^{(4)\,t}$.
However, it is easy to check that the algebra
generated by the entries of $K$ satisfying
eq. (\ref{3i}) coincides with the algebra determined by (\ref{3h})
with  the additional condition $det_qK=0$.
To see it, the following consequences of the eigenvalue decomposition
of $R$ ($\hat{R} \equiv {\cal P}R= qP_+-q^{-1}P_-$) are useful
\begin{equation}\label{3j}
P_-\hat{R}=\hat{R}P_-=-q^{-1}P_- \quad , \quad
P_-\hat{R}^{-1}=\hat{R}^{-1}P_-=-qP_- \quad ,
\end{equation}
\begin{equation}\label{3k}
q^2\hat{R}^{-1}=\hat{R}- (q^3-q^{-1})P_- \quad .
\end{equation}
\noindent
Multiplying  now eq. (\ref{3i}) by $P_-{\cal P}$ from the left and
by ${\cal P}P_-$ from the right and using (\ref{3j}) we get
\begin{equation}\label{3l}
-q^{-1} P_-K_1 R^{(2)}K_2 {\cal P} P_- =-q^{3}P_- {\cal P}K_2 R^{(3)}K_1P_-
\end{equation}
\noindent
as $\hat{R}^{(i)}={\cal P}R^{(i)}$, $R^{(2)}={\cal P}R^{(3)}{\cal P}$
and $K_1={\cal P}K_2{\cal P}$
\begin{equation}\label{3m}
(q^3-q^{-1})P_-K_1\hat{R}^{(3)}K_1P_-=0 \quad.
\end{equation}
\noindent
Thus, (if $q^4 \neq 1$) we obtain that $det_qKP_-
=P_-K_1\hat{R}^{(3)}K_1P_-=0$.

Now , using (\ref{3k}) the RE (\ref{3i}) can be expressed in the following way
\begin{equation}\label{3n}
R_{12} K_{1} R^{(2)} K_{2} =  K_{2} R^{(3)} K_{1} R_{21}
-(q^3-q^{-1})P_-K_2R^{(3)}K_1P_-{\cal P}
\end{equation}
\noindent
and using the definition of the $q$-determinant
\begin{equation}\label{3o}
R_{12} K_{1} R^{(2)} K_{2} =  K_{2} R^{(3)} K_{1} R_{21}
-(q^3-q^{-1}){\cal P} (det_qK) P_-{\cal P}\quad
\end{equation}
\noindent
as $det_qK=0$, we just obtained eq. (\ref{3k}); thus, as the algebra is the
same eq. (\ref{3i}) may be  discarded.

The matrices $R^{(2)}, R^{(3)}$ are not determined, and
characterize the
mixed commutation relations between quantum group elements and conjugated
elements in (\ref{2c}), (\ref{tw2}) and
\begin{equation}\label{3q}
\tilde{T}_{1}^{-1} R^{(2)} T_{2} = T_{2} R^{(2)} \tilde{T}_{1}^{-1} \quad ,
\quad
T_{1} R^{(3)} \tilde{T}_{2}^{-1} = \tilde{T}_{2}^{-1} R^{(3)} T_{1} \; .
\end{equation}

\noindent
Two particularly special cases arise

\vspace{1\baselineskip}

\noindent
{\bf a) Commuting case:} if the two quantum group copies are independent,
the quantum matrices commute
\begin{equation}\label{3r}
T_1\tilde{T}_2=\tilde{T}_2T_1\quad,
\end{equation}
\noindent
here, $R^{(2)} = R^{(3)} =I$, and then,  eq. (\ref{3h}) gives the
reflection equation, which is equivalent to the `$RTT$' relation (\ref{2e})
(see below)
\begin{equation}\label{3s}
R_{12} K_{1} K_{2} = K_{2} K_{1} R_{21}
\end{equation}
\noindent
Eq. (\ref{3i}), in this particular case, produces the RE
\begin{equation}\label{3t}
R_{12} K_{1} K_{2} = q^{2} K_{2} K_{1} R_{12}^{-1}\quad,
\end{equation}
\noindent
however, as we have just shown, this possibility
leads  to the same commutation relations (\ref{3s}) for the entries of $K$
plus  the additional condition $det_{q}K=0$ \cite{AKR}.
The $q$-Minkowski algebra (\ref{3s}) is isomorphic to the quantum group
algebra $GL_q(2)$, by \cite{AKR}
\begin{equation}\label{3u}
T=K\sigma^1  \quad ,\qquad R_{12}T_1T_2=T_2T_1R_{12} \quad,
\end{equation}
\noindent
where $\sigma^1$ is the usual Pauli matrix. Then, it is not possible to
define a linear central element in the algebra generated by the entries
$(\alpha, \beta, \gamma, \delta)$ of matrix
$K$, and the quadratic one is given by the $q$-determinant (\ref{det1})
with $\hat{R}^{(3)}={\cal P}R^{(3)}={\cal P}$
\begin{equation}\label{3v}
det_qK P_-=P_-K_1{\cal P}K_1P_-
= (-q^{-1}) (\alpha \delta -q \gamma \beta ) P_- \quad.
\end{equation}

\vspace{1\baselineskip}

\noindent
{\bf b) Non-commuting case:} now, assuming the non-trivial commutation
relations between the two copies of $SL_q(2,C)$
\begin{equation}\label{3w}
R_{12} T_{1} \tilde{T}_{2} = \tilde{T}_{2} T_{1} R_{12}
\end{equation}

\noindent
we see that (\ref{3q}) is fulfilled for $R^{(2)} = R_{21}$. Then, eq.
(\ref{3k}) leads to the RE
\begin{equation}\label{3x}
R_{12} K_{1} R_{21} K_{2} = K_{2} R_{12} K_{1} R_{21}
\end{equation}

\noindent
Again, (\ref{3i}) produces an equation
\begin{equation}\label{3y}
R_{12} K_{1} R_{21} K_{2} = q^{2} K_{2} R_{12} K_{1} R_{12}^{-1}
\end{equation}
\noindent
which leads to the same commutation relations as (\ref{3x}) with
the restriction $det_{q} K=0$.

These equations \cite{AKR} define the quantum Minkowski algebra of
\cite{WSSW}-\cite{OSWZ},
in which  the linear central term is identified with the time coordinate
and the $q$-determinant, defined by (\ref{det1}) where $\hat{R}^{(3)}
=\hat{R}$
\begin{equation}\label{3z}
det_qK P_-=P_-K_1 \hat{R} K_1P_-
= (-q^{-1}) (\alpha \delta -q^2 \beta \gamma ) P_- \quad,
\end{equation}
\noindent
gives the quadratic central element which is identified with the invariant
$q$-Minkowski length.

\vspace{1\baselineskip}

Having a $q$-vector $X \mapsto TX$ and a $q$-matrix
$K \mapsto TKT^{\dagger}$, it is natural to construct higher rank
tensors transforming as
\begin{equation}\label{ten1}
\varphi : \; L \longmapsto T^{\otimes n} L (T^{\dagger})^{\otimes n}\quad;
\end{equation}
\noindent
they are invariant subspaces of the $q$-Minkowski algebra for the coaction
$ \varphi $. The generators of the $q$-tensors $L$ may be
extracted from matrices of higher dimensions, {\it e.g.}
\begin{equation}\label{ten2}
\begin{array}{c}
L^2 \sim K^{\otimes_q 2}=K_1R_{21}K_2 \quad \longmapsto T_1T_2 K^{\otimes_q2}
T_1^{\dagger}T_2^{\dagger} \quad, \\
\, \\
L^3 \sim K^{\otimes_q 3}=K_1R_{21}K_2 R_{31}R_{32}K_3 \quad , \\
\; \\
L^n \sim K^{\otimes_q n}=K_1 \prod_{j=2}^{n}(R_{j1}R_{j2}...R_{j\,j-1}K_j)
\quad .
\end{array}
\end{equation}
\noindent
These subspaces  (as in the non-deformed theory) are reducible (for instance,
$K^{\otimes_q 2}$
has $det_qK$ as an invariant element). One can apply to $T^{\otimes 2}=T_1T_2$
the appropriate
projector $P^{(1)}$ to  the spin 1 representation (and the same for
$(T^{\dagger})^{\otimes 2}$) to get a tensor of generators transforming
according $D^{1,1}$ {\it irrep} of the $q$-Lorentz group. Quantum tensors
transforming according $D^{j,s}$ {\it irreps} could be  constructed
in the same manner.
We find additional $R$-matrix factors in the tensor products of $K$
(\ref{ten2})
(cf. (\ref{det1})).
This construction is useful for a description of higher spin $q$-wave equations
(see also \cite{pillin}).

\vspace{2\baselineskip}

\noindent
{\bf Acknowledgements:} This research has been partially
sponsored by a CICYT (Spain) research grant. One of us (P.P.K.)
also wishes to thank the DGICYT, Spain, for financial support.

\end{document}